\documentclass[preprintnumbers,superscriptaddress,amsmath,amssymb]{revtex4}
\usepackage{graphicx}
\usepackage{amsmath}\usepackage{slashed}
\usepackage{color}
\usepackage{mathtools}
\usepackage{txfonts}

\UseRawInputEncoding
\begin{document}

\thispagestyle{empty}

\title{Why the Casimir Force for Magnetic Metals Computed by the Lifshitz Theory
Using the Drude Model Disagrees with the Measurement Data}

\author{
G.~L.~Klimchitskaya}
\affiliation{Central Astronomical Observatory at Pulkovo of the Russian Academy of Sciences, St.Petersburg,
196140, Russia}
\affiliation{Peter the Great Saint Petersburg
Polytechnic University, Saint Petersburg, 195251, Russia}

\author{C.~C.~Korikov}
\affiliation{Moscow Center for Advanced Studies, Kulakova Str. 20, 123592 Moscow, Russia}

\author{
V.~M.~Mostepanenko}
\affiliation{Central Astronomical Observatory at Pulkovo of the Russian Academy of Sciences, St.Petersburg,
196140, Russia}
\affiliation{Peter the Great Saint Petersburg
Polytechnic University, Saint Petersburg, 195251, Russia}

\begin{abstract}
{We consider the Casimir force in configurations with magnetic metal plates and
analyze the reasons why the predictions of the Lifshitz theory using the dielectric
permittivity of the Drude model are inconsistent with the measurement data. For this
purpose, the contributions of the electromagnetic waves with the transverse magnetic
and transverse electric polarizations to the Casimir force are computed using the
Lifshitz theory expressed in terms of the pure imaginary Matsubara frequencies.
Furthermore, the fractions of the evanescent and propagating waves in these
contributions are found using an equivalent formulation of the Lifshitz theory
along the real frequency axis. All computations are performed for Au-Ni and
Ni-Ni plates using the Drude model and the experimentally consistent plasma
model over the separation region from 0.5 to 6~$\muup$m, where the
total force value is determined by  conduction
electrons. It is shown that the transverse magnetic contribution to the Casimir force
does not depend on the used model of the dielectric permittivity, so that the total
difference between the predictions of the Lifshitz theory using the Drude model and
the measurement data is determined by the transverse electric contribution. In doing
so, as opposed to the case of nonmagnetic metals, both fractions of the evanescent and
propagating waves in this contribution depend on the model of the dielectric
permittivity used in computations, whereas the magnetic properties of the plate metal
influence the Casimir force solely through the fraction of propagating waves in the
transverse electric contribution. The issue of a more adequate theoretical description of
the electromagnetic response of magnetic metals is discussed.
}\end{abstract}

\maketitle

\section{Introduction}
The first quantum theory by Werner Heisenberg~\cite{1} and Erwin Schr\" {o}dinger
\cite{2}, quantum mechanics, created one hundred years ago laid the groundwork for an
understanding of many physical phenomena which could not be described on the basis
of standard approaches developed in the framework of Newtonian mechanics and
Maxwell's electrodynamics. One of these phenomena is the van der Waals force
playing the crucial role in diverse fields of physics, chemistry, biology, and
nanotechnology~\cite{3}. The nonrelativistic theory of the van der Waals forces
was created by London~\cite{4} in 1930 on the basis of quantum mechanics.
A generalization of the van der Waals force for larger separations between the
interacting bodies was suggested by Casimir~\cite{5} for the case of two ideal
metal planes. The Casimir force arises due to a change in the spectrum of
zero-point oscillations of quantized fields caused by the boundary conditions.
The unified theory of the van der Waals and Casimir forces was created by
Lifshitz~\cite{6,7,8} within the formalism which is essentially the thermal
quantum field theory.

With the advent of high-precision laboratory techniques using the
micromechanical torsional oscillators, the Casimir force acting between the
Au-coated surfaces of a plate and a microsphere was measured by R.S. Decca
with increased precision, and the obtained measurement data were compared
with theoretical predictions of the Lifshitz theory~\cite{9,10,11,12}. The
Lifshitz theory represents the Casimir force as a sum of integrals where the
integrands are expressed via the dielectric permittivities
$\varepsilon_l^{(m)}=\varepsilon^{(m)}(i\xi_l)$ of two interacting bodies,
$m=1,2$, calculated at the pure imaginary Matsubara frequencies
$i\xi_l=2\pi ik_BTl/\hbar$ where $k_B$ is the Boltzmann constant, $T$ is the
temperature and $l=1,2,3,...$ \cite{6,7,8,13}. The values of $\varepsilon_l^{(m)}$
for two Au bodies  $\varepsilon_l^{(1)}=\varepsilon_l^{(2)}=\varepsilon_l^{\rm {Au}}$
are calculated by using the Kramers-Kronig relation expressing $\varepsilon_l^{(m)}$
via  $\rm {Im}$~$\varepsilon^{m}(\omega)$, and the values of  $\varepsilon^{(m)}$ at
real frequencies $\omega$ are obtained from the measured complex indices of
refraction $n^{(m)}(\omega)$ of the interacting bodies (see the tables in Ref.~\cite{14}).

The values of $n^{(m)}(\omega)$  and, thus, $\varepsilon^{(m)}(\omega)$ can be, however,
measured only over a restricted frequency region from $\omega_{\rm {min}}$ to
$\omega_{\rm {max}}$.
{ {For instance, for Au $\hbar\omega_{\rm {min}}=0.125~$eV and
$\hbar\omega_{\rm {max}}=10^4~$eV}}~\cite{14}.
In so doing, $\omega_{\rm {max}}$ is usually sufficiently
large, so that one does not need to know the values
of  $\varepsilon^{(m)}(\omega)$ for $\omega>\omega_{\rm {max}}$ in order
to calculate the Casimir force with the required accuracy. This is not the case for the
region of low frequencies $\omega<\omega_{\rm {min}}$, where one must know
$\varepsilon^{(m)}(\omega)$ for calculating the Casimir force by the Lifshitz theory.
This problem is usually solved by making an extrapolation of
 $\rm {Im}$~$\varepsilon^{(m)}(\omega)$ obtained from the optical data
 { {for the complex index of refraction}}
 to lower
frequencies by means of the dielectric permittivity of the Drude model, which
provides much tested description of the conduction electrons in nonmagnetic
metals.

Surprisingly, it was found that the predictions of the Lifshitz theory calculated in this
way are excluded by the measurement data for the Casimir force at the confidence
level up to 99.9\%~\cite{9,10,11,12}. Of even greater surprise is the fact that if
the optical data related to the core electrons alone are extrapolated to lower
frequencies by means of the dielectric permittivity of the { {dissipationless}}
plasma model, the
predictions of the Lifshitz theory agree closely with the measurement data.
This agreement is reached by disregarding the optical data in the frequency
region $(\omega_{\rm {min}},\omega_1)$, where $\omega_1$ corresponds to
the first absorption band of Au determined by core electrons, and
extrapolating the obtained  $\rm {Im}$~$\varepsilon^{(m)}(\omega)$ to the
interval $\omega<\omega_1$ by zero because the dielectric permittivity
of the plasma model is real. In this case, $\varepsilon^{(m)}(i\xi_l)$ is found
from the Kramers-Kronig relation for functions having the double pole
at zero frequency as it holds for the plasma model (see Refs.~\cite{13,15} for
details).

An extrapolation of the optical data caused by the core electrons to lower
frequencies by means of the plasma model was often considered as unjustified
\cite{16,17,18,19,20} because this model disregards the relaxation properties
of conduction electrons and is physically applicable only at very high
frequencies (for instance, in the region of infrared optics), where the
relaxation properties of conduction electrons do not play any role.
In subsequent experiments performed by U. Mohideen with Au surfaces
using an atomic force microscope~\cite{21,22,23,24} and by R.S. Decca
using a micromechanical torsional oscillator~\cite{25} the role of all possible
background effects was carefully analyzed and excluded. It was confirmed with
certainty that the Lifshitz theory is consistent with the measurement data if
it uses an extrapolation of the optical data by means of the plasma model
and is inconsistent with the same data if the Drude model is used. Furthermore,
it was shown that an entropy of the Casimir interaction computed using the
Lifshitz theory violates the third law of thermodynamics (the Nernst heat theorem)
{ {for metals with perfect cristal lattices}}
if the Drude model is used and is thermodynamically consistent when one uses
the plasma model (see Refs.~\cite{13,15} for a review). This
situation was often called "the Casimir puzzle"~\cite{26,27,28}.

Many attempts to solve the Casimir puzzle have been undertaken on both
the theoretical and experimental sides. Specifically, the more exact models of
surface roughness were suggested~\cite{28a1,28a2,28a3,28a4,28a5}, much
attention was paid to a rigorous generalization of the Lifshitz theory for the
configuration of a sphere above a plate used in experiments~\cite{28b1,28b2,28b3,28b4,28b5,28b6,28b7,28b8,28b9,28b10,28b11,
28b12,28b13,28b14,28b15,28b16,28b17}, the corrections due to surface
patches~\cite{28c1,28c2,28c3} and variations of the optical data~\cite{28d1,28d2,28d3,28d4}
in the measured and calculated Casimir forces were investigated, as well as the role
of spatial dispersion in the region of the anomalous skin effect~\cite{28e1,28e2,28e3}
(see the review~\cite{29} for details). However, the contradiction between experiment and
theory was left in place.

On this point,
an important progress was reached in investigating the frequency region which
determines a difference in the theoretical predictions of the Lifshitz theory
using extrapolations of the optical data of metallic plates by means of the
Drude and plasma models. Although in almost all computations the Lifshitz
formula written in terms of the pure imaginary Matsubara frequencies has
been used, there is a mathematically equivalent formulation along the
real frequency axis.

In Refs.~\cite{30,31} it was shown that a difference between the theoretical predictions
for the Casimir force using the Drude and plasma extrapolations of the optical
data of nonmagnetic metals originates from the modes with the transverse
electric polarization at low $\omega$. This difference was also attributed to the
contribution of Foucault currents~\cite{32,33}. Moreover, at separations larger than
the thermal length, which is equal to 6~$\muup$m at $T=300$~K, it was
demonstrated~\cite{34,35,36} that this difference is fully determined by the
contribution of evanescent waves with the transverse electric polarization (recall that
for the propagating waves $k_{\bot}\leq \omega/c$, where $k_{\bot}$ is the magnitude
of the wave vector projection on the Casimir plates, but for the evanescent ones the
inequality $k_{\bot} > \omega/c$ holds).

Finally, in Ref.~\cite{37}, by using the Lifshitz formula along the real frequency axis,
it was shown that for Au plates within the wide separation region from 0.5 to
4~$\muup$m, where the Casimir force is fully determined by conduction electrons,
the difference in theoretical predictions using the Drude and plasma models results
from the transverse electric evanescent waves. This result was correlated with the
fact that for nonmagnetic metals the Drude model is much-tested only in the
region of propagating waves with any polarization. For the transverse magnetic
evanescent waves, there are also a few tests in the physics of surface plasmons
polaritons~\cite{38} and in the near-field optical microscopy~\cite{39,40}. The
phenomena of total internal reflection and frustrated total internal
reflection~\cite{41,42,43} give the possibility to verify the Drude model for
both polarizations of the evanescent electromagnetic waves, but for $k_{\bot}$
only slightly exceeding $\omega/c$. For this reason, Ref.~\cite{37} concluded
that the experiments~\cite{9,10,11,12, 21,22,23,24,25} on measuring the Casimir
force between Au surfaces invalidate the Drude model in the region of transverse
electric evanescent waves. As to the propagating waves of both polarizations
and the transverse magnetic evanescent waves, in these regions of frequencies
the Drude model can be safely used when calculating the Casimir force between
nonmagnetic metals with no contradiction with the measurement data.

In Refs.~\cite{35,36}, the novel experiment in the field of classical electrodynamics
was proposed which allows to independently check the Drude model by measuring
the lateral component of magnetic field emitted by the oscillating magnetic dipole
and reflected from a nonmagnetic metallic surface. It was shown that this
component is fully determined by the transverse electric evanescent waves. This
experiment has been already performed by U. Mohideen~\cite{44} and demonstrated
a failure of the Drude model for transverse electric evanescent waves.

Measurements of the Casimir force with magnetic materials are of particular value
because in this case the theoretical predictions using the Drude and plasma
extrapolations of the optical data interchange their places. Thus, the gradient of the
Casimir force,
{ {which is the most precisely measured quantity}}
 in the sphere-plate geometry, computed using the Drude model for
two Au bodies is smaller than that computed using the plasma model. On the
contrary, for two bodies made of magnetic metal Ni the gradient of the Casimir force
computed using the Drude model is larger than that found by means of the plasma
model.

In the experiments performed by means of an atomic force microscope, the gradient
of the Casimir force was measured between Au-Ni and Ni-Ni surfaces~\cite{45,46,47}.
It was shown that the theoretical predictions using extrapolations of the optical
data for Au and Ni to low frequencies by the Drude model are excluded by the
measurement data, whereas the predictions using the plasma model are
experimentally consistent. In the seminal differential measurement of the Casimir
force between either Au- or Ni-coated sphere and the rotating disk with periodic
sectors made of Au or Ni performed by R.S. Decca using a micromechanical
torsional oscillator~\cite{48}, the theoretical predictions using the Drude and plasma
model extrapolations differ by up to a factor of 1000. As a result, the extrapolation
using the Drude model was unequivocally ruled out and the extrapolation by the
plasma model was found to be in good agreement with the measurement data.

In this article, we investigate what regions of real frequencies are responsible for a
disagreement between the theoretical predictions for the Casimir force
obtained using the Drude model in configurations with magnetic metals  and
 the measurement data. For this purpose, we compute the transverse magnetic
and transverse electric contributions to the Casimir force for both the evanescent
and propagating waves in configurations of Au-Ni and Ni-Ni plates using
the Lifshitz formula along the real frequency axis. All computations are repeated
using the Drude and plasma models within the separation region from 0.5 to
6~$\muup$m where the core electrons do not contribute to the force value.

According to our results, the transverse magnetic contribution to the Casimir force
does not depend on whether the Drude or the plasma model is used in
computations and on the values of magnetic permeabilities of the plate
materials. However, the transverse electric contribution is highly dependent on
the model of the dielectric permittivity used. In doing so, the parts of the
transverse electric contributions determined by the propagating waves
computed using the Drude or the plasma models fully determine a dependence
of the Casimir force on magnetic properties. Thus, the difference between the
theoretical Casimir forces with magnetic metals when using the Drude and
plasma models originate from the contribution of transverse electric waves.
Keeping in mind that the experimentally consistent theory for both nonmagnetic
and magnetic metals uses the plasma model, one concludes that for magnetic
materials the Drude model in the region of transverse electric waves is not
accurate enough. The question on how the Drude model should be modified in
this region for the case of magnetic materials is discussed.

The article is organized as follows. In Section 2, the transverse magnetic and
transverse electric contributions to the Casimir force between Au-Ni and Ni-Ni
plates are computed. Section 3 briefly presents the Lifshitz formula written
in terms of real frequencies and the concepts of the propagating and evanescent
waves. In Section 4, the fractions due to evanescent and propagating waves
in the transverse magnetic contribution to the Casimir force are found for
Au-Ni and Ni-Ni plates. In Section 5, the same is done for the transverse electric
contribution. Section 6 contains a discussion, and we finish with the conclusions
in Section 7.

\section{Contributions of Transverse Magnetic and Transverse Electric
Polarizations to the Casimir Force for Au-Ni and Ni-Ni Plates in the
Formalism of Imaginary Frequencies}
\newcommand{\ve}{{\varepsilon}}
\newcommand{\kb}{{k_{\bot}}}
\newcommand{\skb}{{k_{\bot}^2}}
\newcommand{\xk}{{(i\xi_l,k_{\bot})}}
\newcommand{\wk}{{(\omega,k_{\bot})}}

We consider the Casimir force acting between two parallel metallic plates
spaced $a$ apart at temperature $T$ in thermal equilibrium with the
environment. In the first case, one plate is made of nonmagnetic metal Au
and another one of magnetic (but not magnetized) metal Ni. In the second case,
both plates are made of Ni. These plates are assumed to be sufficiently thick
in order that they could be considered as the semispaces. When calculating the
Casimir force between good metals, this condition is satisfied for the plates
of more than 100~nm thickness \cite{13}.

According to the Lifshitz formula written in terms of the pure imaginary
Matsubara frequencies, the Casimir force per unit area of the plates can be
presented as the sum of two contributions due to the electromagnetic waves
with transverse magnetic (TM) and transverse electric (TE) polarizations
\begin{equation}
F(a,T)=F_{\rm TM}(a,T)+F_{\rm TE}(a,T).
\label{eq1}
\end{equation}

In the modern notations using the concept of reflection coefficients,
these contributions take the form \cite{13}
\begin{equation}
F_{\rm TM,TE}(a,T)=-\frac{k_BT}{\pi}\sum_{l=0}^{\infty}
\left(1-\frac{\delta_{l0}}{2}\right)\int\limits_{0}^{\infty}\kb\,d\kb q_l
\frac{r_{\rm TM,TE}^{(1)}\xk r_{\rm TM,TE}^{(2)}\xk e^{-2aq_l}}{1-
r_{\rm TM,TE}^{(1)}\xk r_{\rm TM,TE}^{(2)}\xk e^{-2aq_l}}.
\label{eq2}
\end{equation}
\noindent
Here, $k_B$ is the Boltzmann constant, $\delta_{lk}$ is the Kronecker
symbol,
\begin{equation}
\xi_l=\frac{2\pi k_BTl}{\hbar},{\ \ }l=0,\,1,\,2,\,\ldots\,,\qquad
q_l=\left(\skb+\frac{\xi_l^2}{c^2}\right)^{1/2},
\label{eq2a}
\end{equation}
\noindent
and the reflection coefficients for the
TM and TE polarizations are given by
\begin{equation}
r_{\rm TM}^{(m)}\xk=
\frac{\ve^{(m)}(i\xi_l)q_l-p_l^{(m)}}{\ve^{(m)}(i\xi_l)q_l+p_l^{(m)}},
\qquad
r_{\rm TE}^{(m)}\xk=
\frac{\mu^{(m)}(i\xi_l)q_l-p_l^{(m)}}{\mu^{(m)}(i\xi_l)q_l+p_l^{(m)}},
\label{eq3}
\end{equation}
\noindent
where
\begin{equation}
p_l^{(m)}=\left[\skb+\ve^{(m)}(i\xi_l)\mu^{(m)}(i\xi_l)
\frac{\xi_l^2}{c^2}\right]^{1/2},
\label{eq4}
\end{equation}
\noindent
$\ve^{(m)}(\omega)$ and $\mu^{(m)}(\omega)$ are the dielectric permittivity
and magnetic permeability of the first ($m=1$) and second ($m=2$) plates.
In the first case considered below, $\ve^{(1)}(\omega)=\ve^{\rm Au}(\omega)$,
$\mu^{(1)}(\omega)=1$, $\ve^{(2)}(\omega)=\ve^{\rm Ni}(\omega)$ and
$\mu^{(2)}(\omega)=\mu^{\rm Ni}(\omega)$. In the second case
$\ve^{(1)}(\omega)=\ve^{(2)}(\omega)=\ve^{\rm Ni}(\omega)$ and
$\mu^{(1)}(\omega)=\mu^{(2)}(\omega)=\mu^{\rm Ni}(\omega)$.

We performed computations of the Casimir force per unit area (\ref{eq1})
and the transverse magnetic and transverse electric contributions to it
(\ref{eq2}) for the Au-Ni and Ni-Ni plates at $T=300~$K within the
separation region from 0.5 to $6~\muup$m. In this region of relatively
large separations, the bound (core) electrons do not contribute to the
Casimir force and one can use in computations the dielectric permittivities
caused by only the conduction electrons \cite{13}. Below we use the
permittivities of either the Drude or the plasma model at the pure imaginary
Matsubara frequencies
\begin{equation}
\ve_D^{(m)}(i\xi_l)=1+\frac{{\omega_p^{(m)}}^2}{\xi_l
[\xi_l+\gamma^{(m)}(T)]}, \qquad
\ve_p^{(m)}(i\xi_l)=1+\frac{{\omega_p^{(m)}}^2}{\xi_l^2}.
\label{eq5}
\end{equation}

\newcommand{\upmu}{{\muup}}
For the Au-Ni plates we have the value of the plasma frequency
$\hbar\omega_p^{(1)}=9.0~$eV and the relaxation parameter
$\hbar\gamma^{(1)}(T=300\,\mbox{K})=0.035~$eV for Au \cite{49}
and $\hbar\omega_p^{(2)}=4.89~$eV
and  $\hbar\gamma^{(2)}(T=300\,\mbox{K})=0.0436~$eV for Ni \cite{14,50,51}.
For Ni-Ni plates $\hbar\omega_p^{(1)}=\hbar\omega_p^{(2)}=4.89~$eV and
$\hbar\gamma^{(1)}(T=300\,\mbox{K})=\hbar\gamma^{(2)}(T=300\,\mbox{K})=0.0436~$eV.

To perform computations by Eqs.~(\ref{eq2})--(\ref{eq4}), one also needs
information about the magnetic permeability at the pure imaginary Matsubara
frequencies. Nickel falls into the category of soft ferromagnets which do
not possess the spontaneous magnetization. Keeping in mind that the Casimir
force is determined by the fluctuating electromagnetic field which exhibits
the zero mean magnetic field, here we consider what is called the initial
permeability corresponding to $\mbox{\boldmath$H$}=0$.

The static value of this permeability, $\mu^{\rm Ni}(0)$, is preserved
up to the frequency $\omega_1\approx 2\pi\times 10^5~$rad/s and becomes
equal to unity in the region $\omega>\omega_2\approx 6\pi\times 10^9~$rad/s
(see Figure~14 in Ref.~\cite{52} using the data of Refs.~\cite{53,54,55}).
Note that the first Matsubara frequency at $T=300~$K is
$\xi_1\sim 10^{14}~$rad/s, i.e., much larger than the frequency $\omega_2$
where the magnetic permeability of Ni drops to unity. Taking into account
that the same situation holds for all ferromagnetic materials, it was
concluded \cite{56} that the magnetic properties can influence the Casimir
force only through the zero-frequency term of the Lifshitz formula (\ref{eq2}).
Taking into account that the value of the static permeability of Ni is
sample-dependent \cite{57}, below we use the value $\mu^{\rm Ni}(0)=110$
as for the samples used in the experiments \cite{45,46,47} on measuring the
Casimir force with magnetic surfaces. We recall that for the Au-Ni plates
$\mu^{(1)}(i\xi_l)=1$,
$\mu^{(2)}(i\xi_l)=\mu^{\rm Ni}(i\xi_l)$ and for the Ni-Ni plates
$\mu^{(1)}(i\xi_l)=\mu^{(2)}(i\xi_l)=\mu^{\rm Ni}(i\xi_l)$ where
$\mu^{\rm Ni}(i\xi_l)$ is
\begin{equation}
\mu^{\rm Ni}(i\xi_l)=\left\{
\begin{array}{cl}
\mu^{\rm Ni}(0),&l=0,\\
1,&l\geqslant 1.
\end{array}
\right.
\label{eq6}
\end{equation}

\begin{figure}[h]
\vspace*{-9cm}
\centering 
{\hspace*{-1.5cm}\includegraphics[width=8.in]{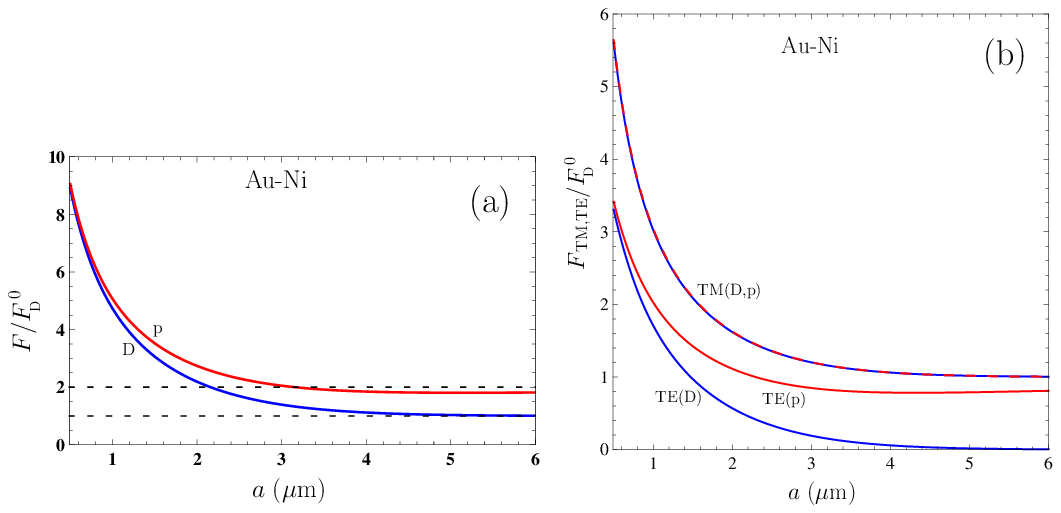}}
\vspace*{-12.cm}
\caption{(a) The Casimir forces per unit area and (b) the transverse magnetic (TM) and
transverse electric (TE) contributions to the Casimir force computed using the Drude (D) and
plasma (p) models in the configuration of Au-Ni plates and normalized to the high-temperature
Casimir force found using the Drude model are shown as the functions of separation.
}
\label{fg1}
\end{figure}

In Figure~\ref{fg1}a, we present the computational results for the total
Casimir forces per unit area computed by Eqs.~(\ref{eq1})--(\ref{eq4})
in the configuration of Au-Ni plates using the Drude and the plasma models
as the functions of separation between plates.
For the sake of convenience in presentation, these results are normalized
to the high-temperature (large separation) limiting case of the Casimir
force between the Au-Ni plates computed by means of the Drude model
\begin{equation}
F_{D}^0(a,T)=-\frac{k_BT}{4\pi a^3}\,\zeta(3),
\label{eq7}
\end{equation}
\noindent
where $\zeta(z)$ is the Riemann zeta function. This is equivalent
to one half of the result for two ideal metal planes \cite{13}.
The upper line is computed using the plasma model and the lower
line --- by the Drude one. It is seen that there is a significant
deviation between these two theoretical predictions.

Next, using Eqs.~(\ref{eq2})--(\ref{eq4}), we computed the contributions
to the total Casimir force between the Au-Ni plates given by the
electromagnetic waves with the TM and TE polarizations. The computational
results for $F_{\rm TM}$ and $F_{\rm TE}$ obtained using the Drude and plasma
models normalized to $F_D^0$ are shown in Figure~\ref{fg1}b as the functions of
separation. The upper line demonstrates the TM contribution to the total
Casimir force which is essentially independent of the used model of the dielectric
permittivity. Note that the TM contribution to the Casimir force does not depend
on the magnetic properties because the TM reflection coefficient at zero Matsubara
frequency $r_{\rm {TM}}(0,k_{\bot})=1$.
By contrast, the TE contributions to the Casimir force
computed using the Drude model (the lower line) and the plasma model (the upper
line) essentially depend on the model used.  As is seen in Figure~\ref{fg1}b,
just the contributions of the TE polarizations computed using the two models determine
the entire difference in the total Casimir forces shown in Figure~\ref{fg1}a.

\begin{figure}[b]
\vspace*{-9.8cm}
\centering 
{\hspace*{-1.5cm}\includegraphics[width=8.in]{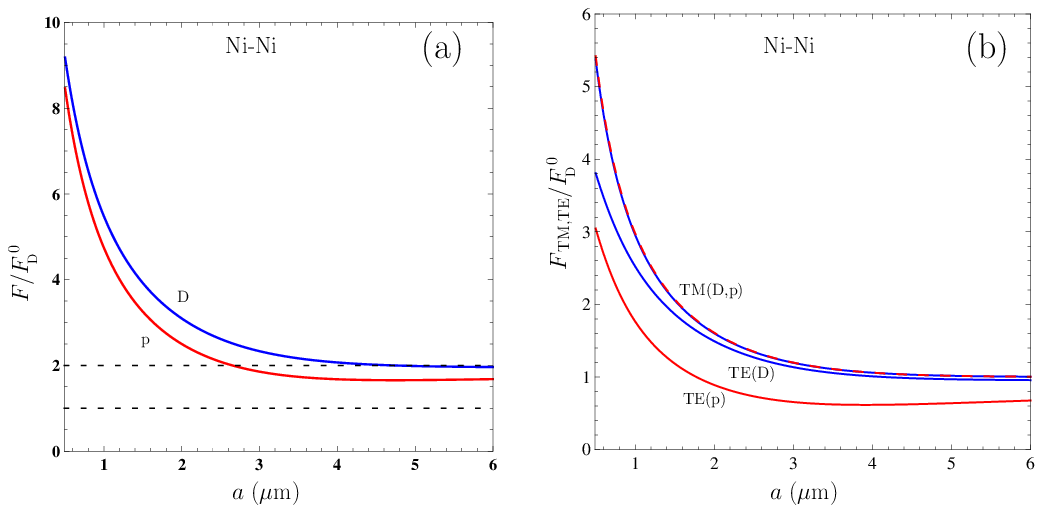}}
\vspace*{-11.7cm}
\caption{(a) The Casimir forces per unit area and (b) the transverse magnetic (TM) and
transverse electric (TE) contributions to the Casimir force computed using the Drude (D) and
plasma (p) models in the configuration of Ni-Ni plates and normalized to the high-temperature
Casimir force found using the Drude model are shown as the functions of separation.
}
\label{fg2}
\end{figure}

Similar results were obtained for the configuration of two magnetic Ni-Ni plates.
In Figure~\ref{fg2}a, the computational results for the total Casimir force
per unit area found using the Drude and plasma models are shown as the functions
of separation between the plates. As in Figure~\ref{fg1}, these results are
normalized to the values of $F_D^0$ defined in Eq.~(\ref{eq7}). According
to Figure~\ref{fg2}a, for two magnetic metals the theoretical predictions
obtained using the Drude and plasma models are more different at short
separations and less different at large separations than for one nonmagnetic
and one magnetic metals. What is more, for two magnetic metals the predictions
of these models interchange their places.

In Figure~\ref{fg2}b, we present the computational results for $F_{\rm {TM}}$ and
$F_{\rm {TE}}$  normalized to $F_D^0$, which are obtained using the Drude and
plasma models as the functions of separation in the configuration of Ni-Ni plates.
The upper line shows the TM contribution to the total Casimir force in this
configuration which is again essentially independent of the used permittivity
model. The TM contribution is also independent of the magnetic properties
of Ni for the reason indicated above. The lower and medium lines show the
contributions of the TE polarization computed using the Drude and plasma models,
respectively. As is seen in Figure~\ref{fg2}b, these predictions differ widely at both
small and large separations. The difference between them fully determines the
difference between the total Casimir forces computed by means of the Drude
and plasma models in Figure~\ref{fg2}a.

According to the above results, the Casimir puzzle for magnetic materials (similar to
nonmagnetic ones~\cite{37}) is caused by the contribution of electromagnetic waves
with the transverse electric polarization. As to the contribution of TM polarized
electromagnetic waves, it can be safely computed using the Drude model with no
contradiction with the measurement data. In order to further bring under control
the roots of the Casimir puzzle for magnetic materials, it is important also to find
out what fractions of the Casimir force in each of the TM and TE contributions are
determined by the propagating and evanescent waves. To accomplish this, it is
necessary to use the formulation of the Lifshitz theory along the real frequency axis.

\section{The Lifshitz Formula in Terms of Real Frequencies for Au-Ni
and Ni-Ni Plates: Propagating and Evanescent Waves}

The Casimir force per unit area of two parallel plates is again represented
by Eq.~(\ref{eq1}) as the sum of the transverse magnetic and transverse
electric contributions. Each of these contributions is now expressed, although
mathematically equivalent to Eq.~(\ref{eq2}), but quite differently as a sum
of fractions determined by the propagating and evanescent waves
\begin{equation}
F_{\rm TM}(a,T)=F_{\rm TM}^{\rm prop}(a,T)+F_{\rm TM}^{\rm evan}(a,T),
\qquad
F_{\rm TE}(a,T)=F_{\rm TE}^{\rm prop}(a,T)+F_{\rm TE}^{\rm evan}(a,T),
\label{eq8}
\end{equation}
\noindent
where \cite{13}
\begin{equation}
F_{\rm TM,TE}^{\rm prop}(a,T)=-\frac{\hbar}{2\pi^2}
\int\limits_{0}^{\infty}d\omega\,\coth\frac{\hbar\omega}{2k_BT}
\int\limits_{0}^{\omega/c}\kb\,d\kb\,{\rm Im}\left[q
\frac{r_{\rm TM,TE}^{(1)}\wk r_{\rm TM,TE}^{(2)}\wk e^{-2aq}}{1-
r_{\rm TM,TE}^{(1)}\wk r_{\rm TM,TE}^{(2)}\wk e^{-2aq}}\right],
\label{eq9}
\end{equation}
\noindent
and
\begin{equation}
F_{\rm TM,TE}^{\rm evan}(a,T)=-\frac{\hbar}{2\pi^2}
\int\limits_{0}^{\infty}d\omega\,\coth\frac{\hbar\omega}{2k_BT}
\int\limits_{\omega/c}^{\infty}\kb\,d\kb q\,{\rm Im}
\frac{r_{\rm TM,TE}^{(1)}\wk r_{\rm TM,TE}^{(2)}\wk e^{-2aq}}{1-
r_{\rm TM,TE}^{(1)}\wk r_{\rm TM,TE}^{(2)}\wk e^{-2aq}}.
\label{eq10}
\end{equation}

The reflection coefficients in Eqs.~(\ref{eq9}) and (\ref{eq10}) are similar
to Eq.~(\ref{eq3})  but depend on real frequencies $\omega$:
\begin{equation}
r_{\rm TM}^{(m)}\wk=
\frac{\ve^{(m)}(\omega)q-p^{(m)}}{\ve^{(m)}(\omega)q+p^{(m)}},
\qquad
r_{\rm TE}^{(m)}\wk=
\frac{\mu^{(m)}(\omega)q-p^{(m)}}{\mu^{(m)}(\omega)q+p^{(m)}},
\label{eq11}
\end{equation}
\noindent
where now
\begin{equation}
q=\left(\skb-\frac{\omega^2}{c^2}\right)^{1/2}, \qquad
p^{(m)}=\left[\skb-\ve^{(m)}(\omega)\mu^{(m)}(\omega)
\frac{\omega^2}{c^2}\right]^{1/2}.
\label{eq12}
\end{equation}

The main feature of the propagating electromagnetic waves, which
determine the fraction $F_{\rm TM,TE}^{\rm prop}$ in Eq.~(\ref{eq9}),
is that $\kb=(k_1^2+k_2^2)^{1/2}\leqslant\omega/c$, i.e., the mass-shell
equation $\skb+k_3^2=\omega^2/c^2$, following from the wave equation,
is satisfied with some real component of the wave vector $k_3$
(we assume that the Casimir plates are in the plane $xy$ and the $z$
axis is perpendicular to it). For the evanescent waves, which
determine the fraction $F_{\rm TM,TE}^{\rm evan}$ in Eq.~(\ref{eq10}),
contrastingly, it holds $\kb>\omega/c$, i.e., the mass-shell equation
is satisfied with only the pure imaginary { {$k_3=iq$}}.
This leads to the exponentially fast decay,
{ {$\sim\exp(ik_3z)=\exp(-qz)$}},
of the evanescent waves in the direction perpendicular to the plate
explaining another name, "the surface waves", for this physical phenomenon.

Equations (\ref{eq8})--(\ref{eq10}) are inconvenient for using in
numerical computations of the Casimir force. The problem is that the
fraction of the propagating waves in Eq.~(\ref{eq9}), in accordance with
Eq.~(\ref{eq12}), contains the exponent of pure imaginary power $-2aq$,
which necessitates to calculate the integrals of the quickly oscillating
functions ({ {see, e.g., Ref.}}~\cite{89a}).
At the same time, for the fractions of the evanescent waves
in Eq.~(\ref{eq10}), the quantity $q$ is real which ensures against quick
oscillations of the integrands and makes possible the numerical integration.

In order to perform computations of the fractions of evanescent and propagating
waves (\ref{eq9}) and (\ref{eq10}), one needs the dielectric permittivities of Au
and Ni plates and the magnetic permeability of the Ni plate along the real frequency
axis. Thus, the dielectric permittivities of the Drude and plasma models
are given by
\begin{equation}
\ve_{\rm D}^{(m)}(\omega)=1-\frac{(\omega_p^{(m)})^2}{\omega[\omega+i\gamma^{(m)}(T)]},
\qquad
\ve_{\rm p}^{(m)}(\omega)=1-\frac{(\omega_p^{(m)})^2}{\omega^2},
\label{eq13}
\end{equation}
\noindent
where the values of the plasma frequencies and relaxation parameters for Au and Ni
were indicated in Section 2.

\begin{figure}[t]
\vspace*{-12cm}
\centering 
{\hspace*{-2.5cm}\includegraphics[width=8.in]{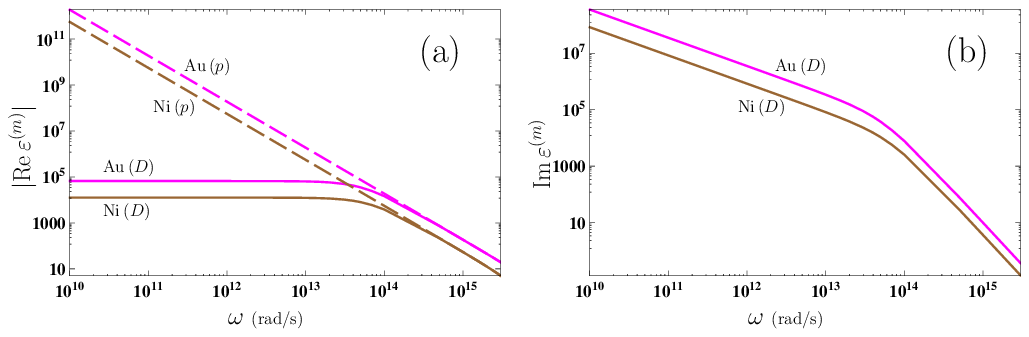}}
\vspace*{-11.8cm}
\caption{(a) The magnitudes of real parts of the dielectric permittivities
of the Drude and plasma models and (b) the imaginary parts of the dielectric
permittivity of the Drude model are shown by the pairs of solid (the Drude model)
and dashed (the plasma model) lines as the functions of frequency.
In all cases the upper and lower lines in each pair are plotted for Au and Ni
plates, respectively.
}
\label{fg3}
\end{figure}

In Figure~\ref{fg3}, the magnitude of the real part of dielectric permittivity
of the Drude model (a) and its imaginary part (b) are shown by the upper and
lower solid lines for Au and Ni plates, respectively.
The upper and lower dashed lines in Figure~\ref{fg3}a show the magnitude of
dielectric permittivity of the plasma model, which is real, for Au and Ni plates,
respectively. Figures~\ref{fg3}a and \ref{fg3}b are plotted up to the largest frequencies
contributing to the computational results over the region of chosen separations from
0.5 to $6~\upmu$m. The plotted straight lines can be extrapolated to the
frequencies below $10^{10}~$rad/s contributing to the Casimir force.

As discussed in Section~2, when using the Lifshitz formula for magnetic materials
written in terms of Matsubara frequencies, at not too low temperature one needs
to know only the value of the static magnetic permeability.  However, the computations
of the Casimir force using Eqs.~(\ref{eq9})--(\ref{eq12}) written in terms of real
frequencies require the knowledge of magnetic permeability over the wide frequency
region. According to discussion in Section~2, the static magnetic permeability
of Ni sample under consideration $\mu(0)=110$ preserves its value up to the
frequency $\omega_1\approx 2\pi\times 10^5~$rad/s and is equal to unity in
the region $\omega>\omega_2\approx 6\pi\times 10^9~$rad/s \cite{51,52,53,54}.
Within the frequency region $(\omega_1,\omega_2)$, where the magnetic permeability
of Ni essentially depends on the frequency, it can be approximately described
by the Debye model \cite{51}. Combining these pieces of information, the magnetic
permeability of Ni over the entire axis of real frequencies can be represented as
\begin{equation}
\mu^{(m)}(\omega)=\left\{
\begin{array}{cl}
\mu^{\rm Ni}(0), & 0\leqslant\omega\leqslant\omega_1, \\[1mm]
1+\frac{\mu^{\rm Ni}(0)-1}{1-i{\omega}/{\omega_{ch}}}, &
\omega_1<\omega\leqslant\omega_2, \\[1mm]
1, &\omega>\omega_2,
\end{array}
\right.
\label{eq14}
\end{equation}
\noindent
where  the characteristic frequency $\omega_{ch}\approx 2\pi\times 10^7~$rad/s
{ {is obtained using Figure~14 in Ref.}}~\cite{52}.
{ {Note that the computational results presented below do not depend on
variations in the value of $\omega_{ch}$ if the values of $\omega_1$ and
$\omega_2$ remain unchanged.}}

\begin{figure}[b]
\vspace*{-12cm}
\centering 
{\hspace*{-1.5cm}\includegraphics[width=8.in]{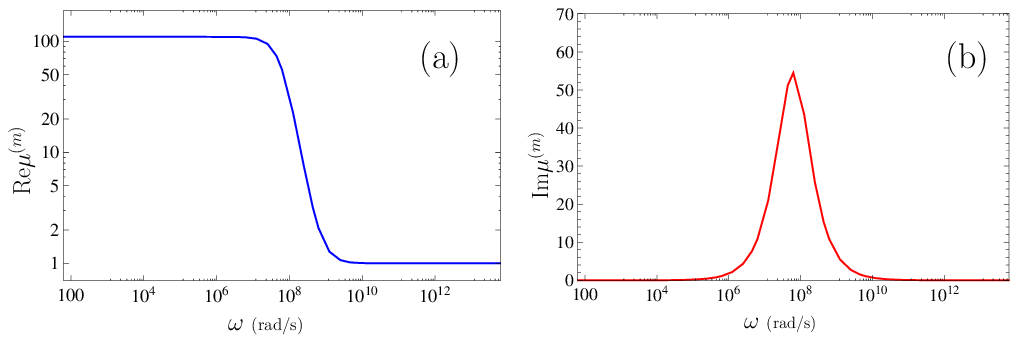}}
\vspace*{-11.8cm}
\caption{(a) The real and (b) imaginary parts of the magnetic permeability
of Ni are shown as the functions of frequency.
}
\label{fg4}
\end{figure}

In Figure~\ref{fg4}, the real (a) and imaginary (b) parts of the magnetic
permeability (\ref{eq14}) are shown as the functions of frequency.
In the next two sections, the data of Figures~\ref{fg3} and \ref{fg4} are used
for computations of the fractions of evanescent and propagating waves in the
transverse magnetic and transverse electric contributions to the Casimir force.

\section{Fractions due to Evanescent and Propagating Waves in the Transverse
Magnetic Contributions to the Casimir Force for Au-Ni and Ni-Ni Plates
in the Formalism of Real Frequencies}

We start from the computation of the fraction of evanescent waves (\ref{eq10})
in the transverse magnetic contribution to the  Casimir force $F_{\rm TM}$
defined in Eq.~({\ref{eq2}). It is convenient to perform numerical computations
by Eq.~(\ref{eq10}) using, instead of the dimensional variables $\omega$
and $\kb$, the dimensionless ones defined as
\begin{equation}
t=\frac{\omega}{\omega_c}\equiv\frac{2a}{c}\omega, \qquad
w=2a\kb-t.
\label{eq15}
\end{equation}

In terms of these variables, the fraction of evanescent waves (\ref{eq10}) in the
TM contribution to the Casimir force takes the form
\begin{equation}
F_{\rm TM}^{\rm evan}(a,T)=-\frac{\hbar c}{32\pi^2a^4}
\int\limits_{0}^{\infty}dt\,\coth\left(\frac{\hbar c}{4ak_BT}t\right)
\int\limits_{0}^{\infty}dw(w+t)\sqrt{w^2+2wt}\,
{\rm Im}
\frac{r_{\rm TM}^{(1)}(t,w) r_{\rm TM}^{(2)}(t,w) e^{-\sqrt{w^2+2wt}}}{1-
r_{\rm TM}^{(1)}(t,w) r_{\rm TM}^{(2)}(t,w) e^{-\sqrt{w^2+2wt}}}.
\label{eq16}
\end{equation}
\noindent
Here, the reflection coefficient expressed via the variables (\ref{eq15}) is given by
\begin{equation}
r_{\rm TM}^{(m)}(t,w)=\frac{\ve^{(m)}(\omega_ct)\sqrt{w^2+2wt}-
\sqrt{(w+t)^2-\ve^{(m)}(\omega_ct)\mu^{(m)}(\omega_ct)t^2}}{\ve^{(m)}(\omega_ct)
\sqrt{w^2+2wt}+\sqrt{(w+t)^2-\ve^{(m)}(\omega_ct)\mu^{(m)}(\omega_ct)t^2}}.
\label{eq17}
\end{equation}

The dielectric permittivities (\ref{eq13}) entering Eq.~({\ref{eq17}) are
expressed as
\begin{equation}
\ve_{\rm D}^{(m)}(\omega_ct)=1-\frac{({\tilde{\omega}}_p^{(m)})^2}{t
[t+i{\tilde{\gamma}}^{(m)}(T)]},
\qquad
\ve_{\rm p}^{(m)}(\omega_ct)=1-\frac{({\tilde{\omega}}_p^{(m)})^2}{t^2},
\label{eq18}
\end{equation}
\noindent
where ${\tilde{\omega}}_p^2=\omega_p^2/\omega_c^2$ and
$\tilde{\gamma}^{(m)}=\gamma^{(m)}/\omega_c$, and the magnetic permeability of Ni
(\ref{eq14}) is
\begin{equation}
\mu^{(m)}(\omega_ct)=\left\{
\begin{array}{cl}
\mu^{\rm Ni}(0), & 0\leqslant t\leqslant\frac{\omega_1}{\omega_c}, \\[1mm]
1+\frac{\mu^{\rm Ni}(0)-1}{1-i{\omega_ct}/{\omega_{ch}}}, &
\frac{\omega_1}{\omega_c}<t\leqslant\frac{\omega_2}{\omega_c}, \\[1mm]
1, &t>\frac{\omega_2}{\omega_c},
\end{array}
\right.
\label{eq19}
\end{equation}

As discussed in Section~3, computations of the fraction of propagating waves
(\ref{eq9}) in the TM contribution to the Casimir force is made difficult
due to a presence of the rapidly oscillating function under the integral.
This fraction can, however, be found in a simpler way using the full
contribution of the transverse magnetic polarization $F_{\rm TM}$ found in
Section~2 using the Lifshitz formula written in terms of imaginary Matsubara
frequencies
\begin{equation}
F_{\rm TM}^{\rm prop}(a,T)=F_{\rm TM}(a,T)-F_{\rm TM}^{\rm evan}(a,T).
\label{eq20}
\end{equation}

In Figure~\ref{fg5}, the lower and upper dashed lines present the computational
results normalized to $F_D^0$ for the fractions of evanescent and propagating waves
in $F_{\rm TM}$ computed in the configuration of Au-Ni plates using the Drude model
by Eqs.~(\ref{eq16})--({\ref{eq19}) and (\ref{eq20}), respectively.
\begin{figure}[t]
\vspace*{-9.8cm}
\centering 
{\hspace*{-2.5cm}\includegraphics[width=6in]{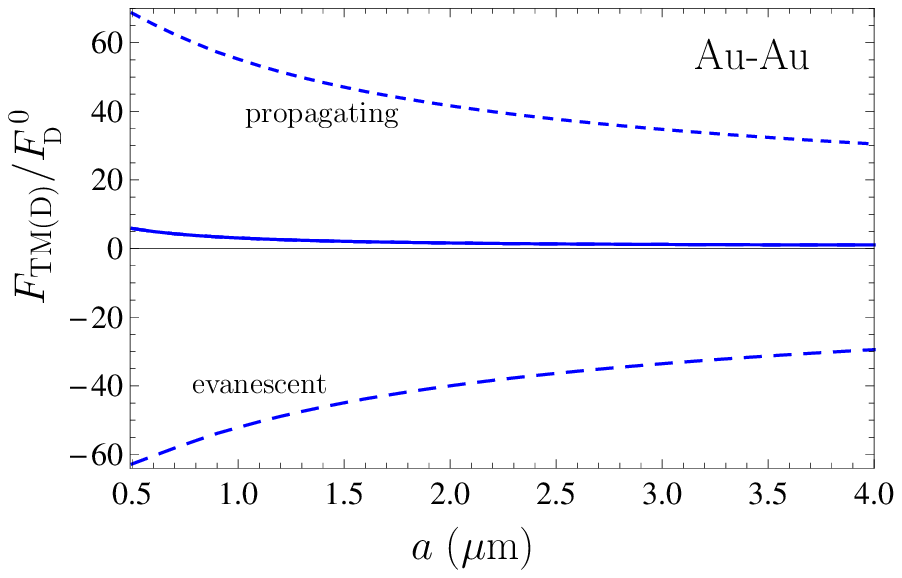}}
\vspace*{-5.3cm}
\caption{The fractions of the evanescent and propagating waves in the transverse
magnetic contribution to the Casimir force per unit area of Au-Ni plates computed
using the Drude model are shown by the lower and upper dashed lines
as the functions of separation.
The total transverse magnetic contribution to the Casimir force computed by
the Drude model shown by the solid line is reproduced from Figure~\ref{fg1}b.
}
\label{fg5}
\end{figure}
The integration in Eq.~(\ref{eq16}) was performed using the adaptive
Gauss-Kronrod quadrature method. It was checked that the obtained results
are stable with respect to changing the number of nodes.
For comparison purposes, the solid line reproduced from Figure~\ref{fg1}b
shows the total transverse magnetic contribution $F_{\rm TM(D)}$.
As is seen in Figure~\ref{fg5}, the contribution of the propagating waves
to the Casimir force is negative [we recall that $F_{\rm D}^0$ defined in
Eq.~(\ref{eq7}) is also negative], which corresponds to attraction, whereas
the contribution of the evanescent waves is positive, which corresponds to
repulsion. Taking into account that
$|F_{\rm TM(D)}^{\rm prop}|>|F_{\rm TM(D)}^{\rm evan}|$, one obtains the
attractive TM contribution to the Casimir force $F_{\rm TM(D)}$ shown by the
solid line in Figure~\ref{fg5}. Note that the fractions of evanescent and
propagating waves in $F_{\rm TM}$ essentially do not depend on the magnetic
properties as in the case with the total $F_{\rm TM}$.

For the plasma model, the dielectric permittivity is real, whereas the imaginary
part of the magnetic permeability of Ni results in only a negligibly small
contribution to $F_{\rm TM}^{\rm evan}$. As a result, if the plasma model is
used in computations, one obtains to  a high degree of precision
\begin{equation}
F_{\rm TM(p)}^{\rm evan}(a,T)=0,\qquad
F_{\rm TM(p)}^{\rm prop}(a,T)=F_{\rm TM(p)}(a,T).
\label{eq21}
\end{equation}

Similar computations have been performed for the configuration of Ni-Ni plates
using the Drude model. The computational results, which differ from the case of
Au-Ni plates only quantitatively, are shown in Figure~\ref{fg6}.
\begin{figure}[b]
\vspace*{-9.3cm}
\centering 
{\hspace*{-2.5cm}\includegraphics[width=6.in]{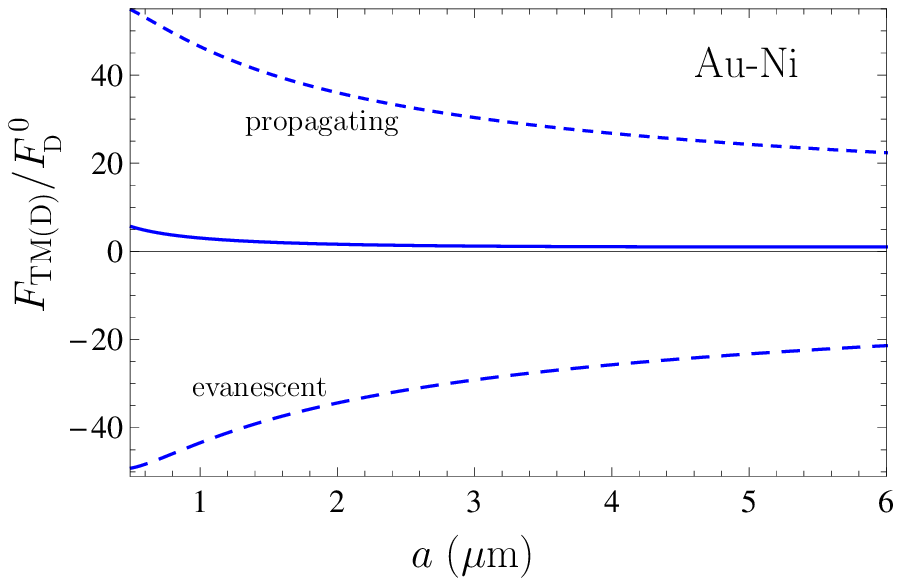}}
\vspace*{-4.9cm}
\caption{The fractions of the evanescent and propagating waves in the transverse
magnetic contribution to the Casimir force per unit area of Ni-Ni plates computed
using the Drude model are shown by the lower and upper dashed lines
as the functions of separation.
The total transverse magnetic contribution to the Casimir force computed by
the Drude model shown by the solid line is reproduced from Figure~\ref{fg2}b.
}
\label{fg6}
\end{figure}
The lower and upper dashed lines demonstrate the fractions of evanescent and
propagating waves in $F_{\rm TM(D)}$ computed using Eqs.~(\ref{eq16})--(\ref{eq19})
and (\ref{eq20}), respectively. The propagating waves again correspond to the
Casimir attraction and the evanescent waves --- to a repulsion. Their sum in
accordance with Eq.~(\ref{eq8}) results in a smaller magnitude attractive TM
contribution to the Casimir force $F_{\rm TM(D)}$ shown in Figure~\ref{fg6} by
the solid line. If the plasma model is used in computations of the fractions of evanescent
and propagating waves in $F_{\rm TM}$ for Ni-Ni plates, one again arrives to
Eq.~(\ref{eq21}), i.e., to a high degree of precision, the evanescent waves do
not contribute to the transverse magnetic part of the Casimir force $F_{\rm TM}$.

Thus, although the transverse magnetic contribution to the Casimir force with
magnetic materials does not depend on whether the Drude or the plasma model is used
in computations, the fractions of evanescent and propagating waves in these
contributions heavily depend on the used permittivity model. Similar situation
was revealed earlier for the case of Casimir force between the nonmagnetic
Au-Au plates~\cite{37}.  Note, however, that due to an error in the computer program
the magnitudes of the fractions of evanescent and propagating waves contributing
to $F_{\rm TM(D)}$ in Figure~4 of Ref.~\cite{37} are shown with much smaller
magnitudes than they really have. The correct fractions of the evanescent and
propagating waves in the transverse magnetic contribution to the Casimir force
for Au-Au plates computed by Eqs.~(\ref{eq16})--(\ref{eq20}) with
$\ve_{\rm D}^{(1)}(\omega_ct)=\ve_{\rm D}^{(2)}(\omega_ct)=
\ve_{\rm D}^{\rm Au}(\omega_ct)$ are shown in Figure~\ref{fg7} by the lower and
upper dashed lines as the functions of separation. The solid line indicates the
total TM contribution $F_{\rm TM(D)}$ to the Casimir force coinciding with that
shown in Figure~4 of Ref.~\cite{37}.

\begin{figure}[t]
\vspace*{-9.3cm}
\centering 
{\hspace*{-2.5cm}\includegraphics[width=6.in]{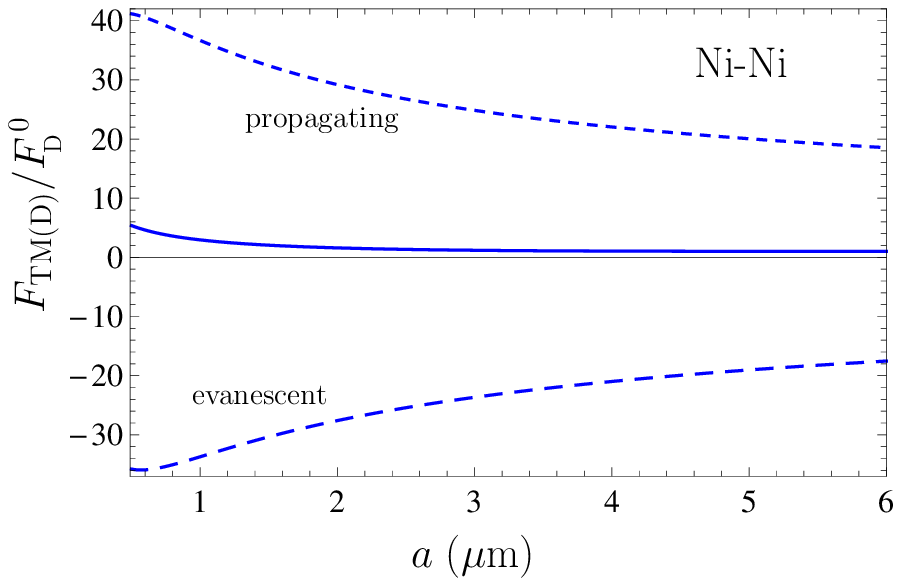}}
\vspace*{-4.9cm}
\caption{The fractions of the evanescent and propagating waves in the transverse
magnetic contribution to the Casimir force per unit area of Au-Au plates computed
using the Drude model are shown by the lower and upper dashed lines
as the functions of separation.
}
\label{fg7}
\end{figure}

\section{Fractions due to Evanescent and Propagating Waves in the Transverse
Electric Contributions to the Casimir Force for Au-Ni and Ni-Ni Plates
in the Formalism of Real Frequencies}

We are coming now to the constituent parts of the transverse electric contribution to
the Casimir force with magnetic plates which are of major importance in the context
of the Casimir puzzle.
In terms of the dimensionless variables (\ref{eq15}), the fraction of evanescent
waves in the TE contribution (\ref{eq10}) is expressed similar to Eq.~(\ref{eq16})
\begin{equation}
F_{\rm TE}^{\rm evan}(a,T)=-\frac{\hbar c}{32\pi^2a^4}
\int\limits_{0}^{\infty}dt\,\coth\left(\frac{\hbar c}{4ak_BT}t\right)
\int\limits_{0}^{\infty}dw(w+t)\sqrt{w^2+2wt}\,
{\rm Im}
\frac{r_{\rm TE}^{(1)}(t,w) r_{\rm TE}^{(2)}(t,w) e^{-\sqrt{w^2+2wt}}}{1-
r_{\rm TE}^{(1)}(t,w) r_{\rm TE}^{(2)}(t,w) e^{-\sqrt{w^2+2wt}}}.
\label{eq22}
\end{equation}
\noindent
The TE reflection coefficient expressed via the variables (\ref{eq15}) is given by
\begin{equation}
r_{\rm TE}^{(m)}(t,w)=\frac{\mu^{(m)}(\omega_ct)\sqrt{w^2+2wt}-
\sqrt{(w+t)^2-\ve^{(m)}(\omega_ct)\mu^{(m)}(\omega_ct)t^2}}{\mu^{(m)}(\omega_ct)
\sqrt{w^2+2wt}+\sqrt{(w+t)^2-\ve^{(m)}(\omega_ct)\mu^{(m)}(\omega_ct)t^2}},
\label{eq22a}
\end{equation}
\noindent
where the dielectric permittivities and magnetic permeability are given by
Eqs.~(\ref{eq18}) and (\ref{eq19}).

In order to avoid a challenging task of integration the rapidly oscillating
functions in Eq.~(\ref{eq9}), the fraction of the propagating waves is found
from
\begin{equation}
F_{\rm TE}^{\rm prop}(a,T)=F_{\rm TE}(a,T)-F_{\rm TE}^{\rm evan}(a,T),
\label{eq23}
\end{equation}
\noindent
where $F_{\rm TE}$ is computed in Section~2 using the formalism of pure imaginary
Matsubara frequencies.

The computational results for the fractions of evanescent and propagating waves
obtained by Eqs.~(\ref{eq22})--(\ref{eq23}) using the Drude model by means
of the adaptive Gauss-Kronrod quadrature method for the case of Au-Ni plates
normalized to $F_D^0$
are shown by the lower and upper dashed lines in Figure~\ref{fg8} as the functions
of separation. The solid line reproduced from Figure~\ref{fg1}b for comparison
purposes shows the total transverse electric contribution to the Casimir force
for Au-Ni plates computed using the Drude model. As is seen in Figure~\ref{fg8},
the propagating waves are again responsible for attraction and the evanescent
ones for repulsion. Taking into account that the attractive part is larger in
magnitude, the total transverse electric contribution is attractive similar to
the transverse magnetic one (see the solid line in Figure~\ref{fg8}).
According to our computational results, the fraction of evanescent waves
$F_{\rm TE(D)}^{\rm evan}$ does not depend on magnetic properties to a high
degree of precision.
\begin{figure}[t]
\vspace*{-9.8cm}
\centering 
{\hspace*{-2.5cm}\includegraphics[width=6.in]{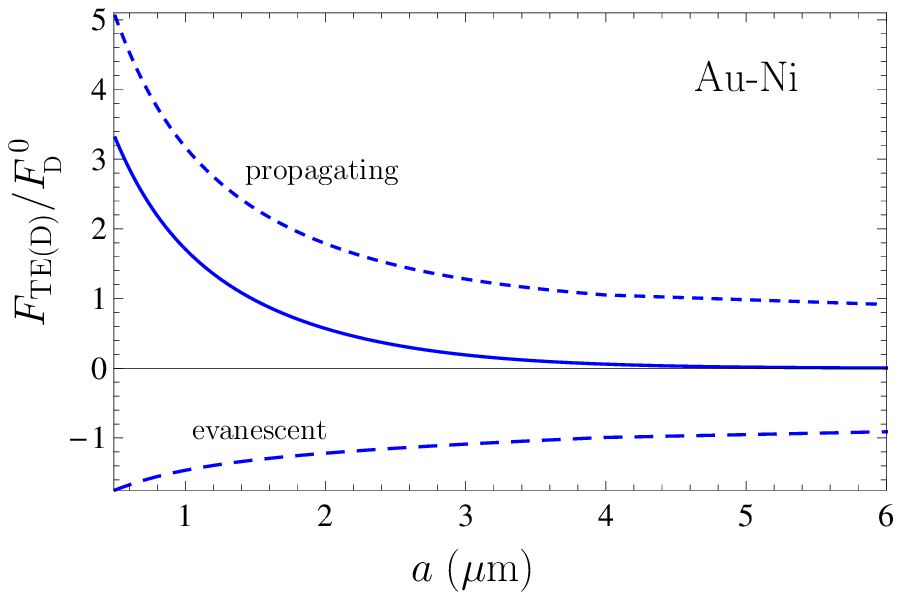}}
\vspace*{-5.3cm}
\caption{The fractions of the evanescent and propagating waves in the transverse
electric contribution to the Casimir force per unit area of Au-Ni plates computed
using the Drude model are shown by the lower and upper dashed lines
as the functions of separation.
The total transverse electric contribution to the Casimir force computed by
the Drude model shown by the solid line is reproduced from Figure~\ref{fg1}b.
}
\label{fg8}
\end{figure}

Now, computations of the fractions of evanescent and propagating waves in
the transverse electric contribution to the Casimir force for Au-Ni plates
by Eqs.~(\ref{eq22})--(\ref{eq23}) are performed using the plasma model.
The imaginary part of the magnetic permeability (\ref{eq19}) leads to
a negligibly small contribution to $F_{\rm TE}^{\rm evan}$. As a result,
from Eq.~(\ref{eq23}) we have
\begin{equation}
F_{\rm TE(p)}^{\rm evan}(a,T)=0,\qquad
F_{\rm TE(p)}^{\rm prop}(a,T)=F_{\rm TE(p)}(a,T).
\label{eq24}
\end{equation}
\noindent
Thus, $F_{\rm TE(p)}^{\rm evan}$ also does not depend on the magnetic
properties of Ni. In fact for Au-Ni plates the Casimir force depends
on magnetic properties only if the plasma model is used in computations
and this dependence is contained in $F_{\rm TE(p)}^{\rm prop}$.

Computation of the fractions of evanescent and propagating waves by
Eqs.~(\ref{eq22})--(\ref{eq23}) using the Drude and plasma models
was also performed for the case of Ni-Ni plates. The computational
results using the Drude model normalized to $F_D^0$ are shown in Figure~\ref{fg9}
where the lower and upper dashed lines show the fractions of evanescent
and propagating waves, respectively, whereas the solid line reproduced
from Figure~\ref{fg2}b represents the total transverse electric contribution.
Its part determined by the propagating waves is larger in magnitude, so
that the transverse electric contribution to the force is attractive.
Similar to the case of Au-Ni plates, for Ni-Ni plates
$F_{\rm TE(D)}^{\rm evan}$ is practically independent of magnetic properties.
\begin{figure}[b]
\vspace*{-9.6cm}
\centering 
{\hspace*{-2.5cm}\includegraphics[width=6.in]{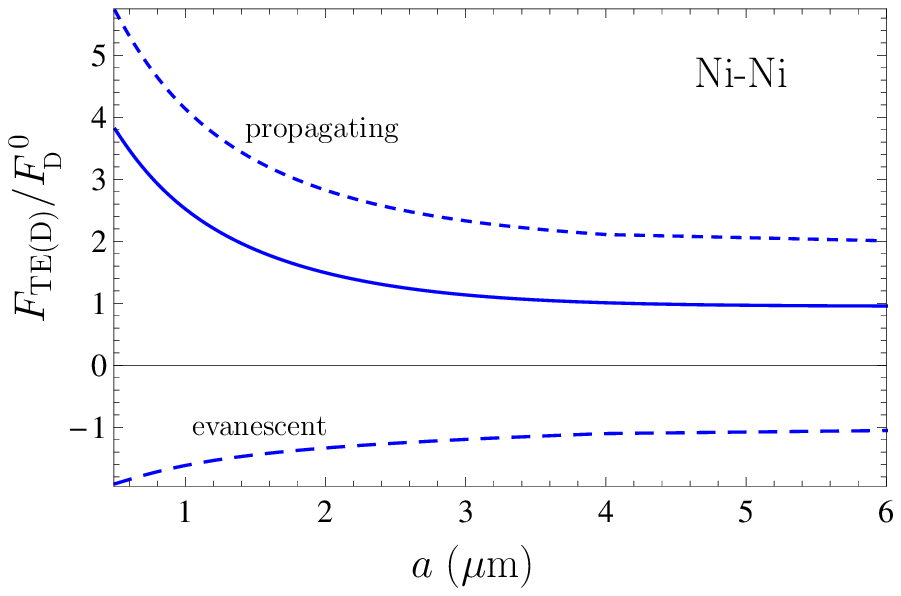}}
\vspace*{-5.3cm}
\caption{The fractions of the evanescent and propagating waves in the transverse
electric contribution to the Casimir force per unit area of Ni-Ni plates computed
using the Drude model are shown by the lower and upper dashed lines
as the functions of separation.
The total transverse electric contribution to the Casimir force computed by
the Drude model shown by the solid line is reproduced from Figure~\ref{fg2}b.
}
\label{fg9}
\end{figure}

If the plasma model is used in computations for the case of Ni-Ni plates,
one again arrives at Eq.~(\ref{eq24}), i.e., to the fact that the fraction
of evanescent waves $F_{\rm TE(p)}^{\rm evan}$ is negligibly small.
Similar to the case of Au-Ni plates, for Ni-Ni plates $F_{\rm TE(D,p)}^{\rm evan}$
are practically independent on magnetic properties. The dependence of magnetic
properties is contained in $F_{\rm TE}^{\rm prop}$ irrespective of whether the
Drude or the plasma model is used.

From the above results, it is seen that there are both similarities and
distinctions between the cases of nonmagnetic and magnetic metals as to
the origin of the Casimir puzzle. According to Section~1, for nonmagnetic
metals the difference in theoretical Casimir forces calculated using the
Drude and plasma models is due to the fraction of transverse electric
evanescent waves. In doing so, the total contribution of transverse
magnetic waves and also the fraction of propagating waves in the contribution
of transverse electric polarization to the Casimir force can be calculated
using the Drude model with no contradiction with the measurement data.
In a similar way, for magnetic metals, the total contribution of transverse
magnetic polarization does not depend on whether the Drude or the plasma model
is used.

However, as opposed to the case of nonmagnetic metals,
for magnetic metals not only the fraction of
evanescent waves, but also the fraction of propagating waves in the transverse
electric contribution to the Casimir force as well depends on the model of
dielectric permittivity used. To see this, one should compare the upper dashed
line in Figure~\ref{fg8} representing $F_{\rm TE(D)}^{\rm prop}$ with  the
middle line in Figure~\ref{fg1}b representing
$F_{\rm TE(p)}=F_{\rm TE(p)}^{\rm prop}$ for the Au-Ni plates and
the upper dashed
line in Figure~\ref{fg9} representing $F_{\rm TE(D)}^{\rm prop}$ with  the
bottom line in Figure~\ref{fg2}b representing
$F_{\rm TE(p)}=F_{\rm TE(p)}^{\rm prop}$ for the Ni-Ni plates.
From this it follows that for the Casimir force with magnetic plates
a contradiction between computations by means of the Lifshitz theory using
the Drude model and the measurement data is due to the contribution of
transverse electric polarization.

It is worth noting that, according to the results of Section~2, the transverse
magnetic contribution to the Casimir force between metallic plates does not
depend on the value of magnetic permeability. In addition, as shown in this
section, the fraction of evanescent waves in the transverse electric contribution
to the Casimir force with magnetic plates also does not depend on the
magnetic permeability with pinpoint accuracy for both the Drude and the plasma
models. In doing so, $F_{\rm TE(p)}^{\rm evan}=0$ to a high precision whereas
$F_{\rm TE(D)}^{\rm evan}$, although is not equal to zero, is independent
of $\mu$.
{ {An independence on $\mu$ in the above mentioned cases arises due to the
fact that $\mu$ is not equal to unity only in the frequency region lying much below
the characteristic interval around the frequency $c/(2a)$ giving the major
contribution to the Casimir force.}}
Thus, an impact of magnetic properties on the Casimir force between
metallic plates is determined by only a fraction of the propagating waves in
the transverse electric contribution.

\section{Discussion}

The Casimir force in configurations with magnetic plates considered in this article
is the subject of prime importance because in this case a comparison between
experiment and theory provides the most convincing and unambiguous
evidence concerning the Casimir puzzle and places strong constraints on the ways of its
resolution. As discussed in Section 1, for the case of nonmagnetic plates the Casimir
puzzle is due to a contribution of transverse electric evanescent waves which is
incorrectly described by the Drude model. The inadequacy of the Drude model
in this case was directly demonstrated in the independent experiment~\cite{44}
related to classical electrodynamics.
{ {The use of the plasma model leads to even greater disagreement with this
experiment. The reason is that the characteristic values of the parameter
$ck_{\bot}/\omega$ in the experimental configuration of Ref}}.~\cite{44}
{ {and in
measurements of the Casimir force differ by 5 orders of magnitude.}}

According to the above results, for the Casimir force with magnetic plates the
situation is more complicated. We have shown that in this case the difference in
theoretical predictions of the Lifshitz theory using the Drude and plasma models
 resulting in contradiction with the measurement data originates from the transverse
electric contribution given by both the evanescent and propagating waves.
In doing so, only a fraction of the propagating waves with the transverse electric
polarization depends on the magnetic properties of the plates. This raises a
question of whether one could doubt in the applicability of the Drude model to
magnetic plates not only in the region of the TE evanescent, but TE propagating
waves as well.

When answering this question positively, it should be remembered that the Drude
model was originally created for the case of nonmagnetic metals and it is not
directly applicable to the ferromagnetic materials (see both the theoretical and
experimental results on this subject in Refs.~\cite{58,59,60,61,62,63,64}).
{ {Unlike the case of nonmagnetic metals, the Drude model fails to correctly
describe the electric conductivity even in the region of propagating waves.
This affects the low-frequency behavior of the TE reflection coefficient.}}
Calculations show that when the Casimir force with magnetic plates is computed
using the Drude model the contribution of TE polarized propagating waves, which
depends on the magnetic permeability, is much larger than in the case when the
plasma model is used in computations. The discussed problems make it necessary
to search for a more adequate response functions of nonmagnetic materials
in the region of transverse electric evanescent waves and of magnetic
materials for both the evanescent and propagating waves with the transverse
electric polarization.

An experience of graphene, where the response functions are found starting
from the first principles of thermal quantum field theory, suggests that a
generalization of the dielectric permittivity of the Drude model should be spatially
nonlocal and possess the double pole at zero frequency~\cite{65,66}. The
phenomenological examples of the permittivities of such kind were considered for
both nonmagnetic~\cite{67} and magnetic~\cite{68} metals. These permittivities
bring the predictions of the Lifshitz theory in agreement with the measurement
data of all performed experiments on measuring the Casimir force~\cite{69} but
are lacking the necessary fundamental justification. There are also other
nonlocal generalizations of the Drude model motivated by the Casimir
puzzle~\cite{20,70,71}, but the Lifshitz theory using these generalizations is not
consistent with experiment.

It would be interesting to directly check the Drude model for magnetic materials in
the region of transverse electric propagating waves in an experiment similar to that
proposed in Refs.~\cite{35,36} for the case of nonmagnetic metals and already
performed in Ref.~\cite{44} confirming a violation of the Drude model in the
region of TE evanescent waves. The realization of such an experiment would finally
give an insight into the roots of the Casimir puzzle for magnetic materials.

\section{Conclusions}

In the foregoing, we investigated the reasons of disagreement between the predictions
of the Lifshitz theory for the Casimir force between Au-Ni and Ni-Ni plates obtained
using the Drude model and the measurement data of several high-precision
experiments. For this purpose, we have computed not only the total Casimir force in the
mentioned configurations and the contributions to it given by the electromagnetic
waves with the TM and TE polarizations, but also the fractions of each of these
contributions given by the evanescent and propagating waves. All computations
were performed using the dielectric permittivities of the Drude and plasma models
and the magnetic permeability of Ni given by the Debye model in the separation
region from 0.5 to 6~$\upmu$m, where the response of metals to the electromagnetic
field is fully determined by conduction electrons.

According to our results, the predictions of the Lifshitz theory for the transverse
magnetic contributions to the Casimir force for both Au-Ni and Ni-Ni plates do not
depend on whether the Drude or the plasma model is used in computations. On the
contrary, the transverse electric contribution to the force was found to be highly
dependent on the used model of dielectric permittivity. Just a difference
between the values of the transverse electric contribution computed using the
Drude model and the experimentally consistent plasma model determines a
discrepancy between the predicted and experimental values of the Casimir force.
At the same time, it was shown that the transverse magnetic contribution to the
Casimir force does not depend on the magnetic permeability of the plates
made of magnetic metal.

Using a formulation of the Lifshitz theory along the real frequency axis, we computed
 the fractions of the evanescent and propagating waves in both the transverse magnetic
and transverse electric contributions to the Casimir force. It was shown that the
propagating waves contribute to an attraction and the evanescent ones - to a repulsion
resulting in total in the attractive Casimir force. Both the fractions of evanescent and
propagating waves in the transverse electric contribution to the Casimir force were
found dependent on a model of dielectric permittivity used. This is different
from the case of nonmagnetic metals investigated earlier where only the fraction
of evanescent waves in the transverse electric contribution to the Casimir force
depends on whether the Drude or the plasma model is used in computations. On the
other hand, it was shown that the impact of magnetic properties is caused by only
a fraction of the propagating waves in the transverse electric contribution to the
Casimir force.

To conclude, for magnetic metals the Drude model fails to precisely describe the
electromagnetic response not only in the region of transverse electric evanescent
waves (for nonmagnetic metals this was already independently confirmed experimentally),
but in the region of transverse electric propagating waves as well. This allows to make
primary emphasis upon the search for a more adequate theoretical description of
the electromagnetic response of magnetic metals on the basis of first physical
principles.

The work of G.L.K. and V.M.M. was partially
supported by the State Assignment for Basic Research (project FSEG-2026-0018). 


\end{document}